\documentclass[a4paper,11pt]{article}
\usepackage[pdftex]{graphicx}

\usepackage{amsmath}
\usepackage{amssymb}
\usepackage{cite} %\renewcommand\citeleft{(} \renewcommand\citeright{)}  
\usepackage{url}
\newcommand{\pPower}{p_{\text{power}}(t)}

\setkeys{Gin}{width=8.7cm, height=6.15cm}
\begin{document}

\title{Evidence for super-exponentially accelerating atmospheric carbon dioxide growth}
%Exponential and Super-Exponential Population and atmospheric CO2 Dynamics: \\No Improvements on the Horizon
\author{%
A.D. H\"usler and D. Sornette \\
Department of Management, Technology and Economics, \\
ETH Zurich, Kreuzplatz 5, CH-8032 Zurich, Switzerland
}
%\date{This version: \today}
\date{This version: 2011-03-17}

\maketitle

\begin{abstract}
We analyze the growth rates of atmospheric carbon dioxide and human population, by comparing the relative merits of two benchmark models, the exponential law and the finite-time-singular (FTS) power law. The later results from positive feedbacks, either direct or mediated by other dynamical variables, as shown in our presentation of a simple endogenous macroeconomic dynamical growth model. Our empirical calibrations finds that the human population 
has decelerated from its previous super-exponential growth until 1960 to a slower-than-exponential growth associated with a
decreasing growth rate. However, the past decade is found to be characterized
by an almost stable growth rate approximately equal to $r(2010) \approx 1.0\%$ per year, suggesting
that the population growth is stabilizing at  ``just'' an exponential growth. As for atmospheric CO2 content, we find that it is at least exponentially increasing and most likely characterized by an accelerating growth rate as off 2009, consistent with an unsustainable FTS power law regime announcing a drastic change of regime. The coexistence of a quasi-exponential growth of human population with a super-exponential growth of carbon dioxide content in the atmosphere is a diagnostic that, until now, improvements in carbon efficiency per unit of production worldwide has been dramatically insufficient.
\end{abstract}

%\keywords{endogenous growth | macroeconomic model | power-law}

%% optional
%\abbreviations{$\textrm{CO}_2$, carbon dioxide; FTS, finite time singularity}
%%%%%%%%%%%%%%%%%%%%%%%%%%%%%%%%%%%%%%%%%%%%%%%%%%%%%%%%%%%%%%%%%%%%%%%%%%%%%%%

\section{Introduction}

\emph{Today humanity uses the equivalent of 1.5 planets to provide the resources we use and absorb our waste. This means it now takes the Earth one year and six months to regenerate what we use in a year.\footnote{\url{http://www.footprintnetwork.org/en/index.php/GFN/page/world_footprint/}}}
%\end{quote}
--- Is humanity inevitably doomed?

During the 1960, leaders were most concerned about human population growth (see for instance \cite{vonFoerster1960}) and about depletion of energy resources (see for example the first report by the Club of Rome \cite{Meadows1972} and its recent reassessment \cite{HallDay}). As a matter of fact, the growth rate\footnote{The growth rate $r$ of the human population (or of any other variable) is defined by expression (\ref{eq:malthus}).  Thus, a constant growth rate corresponds to a  population growing exponentially, with a doubling time given by $(\log 2)/r$.  As the present growth rate is $r(2010) \approx 1.0\%$ per year, this gives a present doubling time of 69 years. If nothing changes, the present 6.8 billion people will be more than 13 billion in 2079. This is in contradiction with projections of OECD for instance and other international organizations, which optimistically expect human population to stabilize around 9 billion individuals.} of human population has peaked in the late 1960s and although population is still growing, it is no longer the prime concern of policy leaders.
This  may be ill-advised as we show below that population growth is not decelerating anymore, being on a
exponential (proportional) growth trajectory.

More recently, scientist and politicians became aware of global warming (see \cite{Weart2008} for a historic overview), 
perhaps due to or augmented by anthropogenic effects (we do not enter this debate). We focus here on the undisputed fact that,
due to the massive use of fossil energies, the world economy emits, among
many other products, large amounts of carbon dioxide into the atmosphere. 
Part of this carbon dioxide is later absorbed by the oceans and plants. The fraction of carbon dioxide 
found in the atmosphere is currently around 50\% of the total anthropogenic emissions, with a slight upward trend \cite{Raupach2008}. Once in the atmosphere,
this $\text{CO}_2$ is thought to play a pivotal role in global warming. 
In a recent Nature issue \cite{Rockstrom2009}, climate change due to $\text{CO}_2$ emissions is identified as one of the most pressing problems that mankind needs to address.

Ref.~\cite{Waggoner2002} discuss the IPAT identity which identifies the most important factors which drive carbon dioxide emissions. They write carbon dioxide emissions as the result of three factors
\begin{equation}
\label{eq:IPAT}
I = P \cdot A' \cdot  T,
\end{equation}
where $I$ (impact) denotes the carbon dioxide emissions, $P$ is human population, $A'$  
represents the affluence (measured as gross world product per capita) and $T$ is technology.

The IPAT identity is useful to help thinking about the contributions of different variables and has been
extensively used and discussed in the literature (see for instance \cite{Chertow2000}). 
However, because one deals fundamentally
with a complex dynamical system driven by entangled feedback loops with delays, the IPAT identity falls short, in our opinion,
of providing the framework to understand the inter-relationships between the dynamical variables.
It is especially important to develop a dynamical framework with delays, when studying the time-evolution of
global variables such as atmospheric carbon dioxide content and human population.
Therefore, motivated by a dynamical view of the human-Earth system, we present here 
a framework borrowing from the theory of endogenous macroeconomic growth \cite{Kremer93,Romer2000}, whose
feedback loops are shown to generate robust regimes of super-exponential growth.
Mathematically, these regimes can be described by simple equations, whose solutions exhibit
finite-time singular (FTS) power law behaviors.  The interest in such solutions is that they
point to change of regimes \cite{Sornette2001, SamSor02, GluzSor02}.

The article is organized as follows.
%Section 2
We presents a simple mathematical framework 
to model growth, first for a single variable like population in the presence of positive feedback, 
and then with several coupled variables, such as population, capital and technology. 
Two benchmark models, the exponential law and the FTS power law, are obtained as limiting cases
of the theoretical framework.
% Section 3
Then we describe the results of the calibration of these two models to some of the most
extensive data sources on human population and atmospheric $\text{CO}_2$ content in the last two centuries
up to present. 
%Section 4 
The final section concludes.

%%%%%%%%%%%%%%%%%%%%%%%%%%%%%%%%%%%%%%%%%%%%%%%%%%%%%%%%%%%%%%%%%%%%%%%%%%%%%%%
\section{Growth models} 

%%%%%%%%%%%%%%%%%%%%%%%%%%%%%%%%%%%%%%%
\subsection{Generalization of Exponential Growth}

The benchmark for population growth is the Malthus model, which postulates that population growth is 
proportional to the population itself, capturing the simple idea that the number of children is
proportional to the number of parents:
\begin{equation}
\label{eq:malthus}
\frac{d p}{d t} = r \cdot p(t)~.
\end{equation} 
The solution of equation (\ref{eq:malthus}) is the exponential function
\begin{equation}
p(t) = a' \exp(r \cdot t)+c'~.
\label{eq:expsol}
\end{equation}

Historically, equation (\ref{eq:malthus}) has been improved by  \cite{Verhulst1845,Verhulst1847} into the logistic equation,
to account for the competition for scarce resources between individuals.
This competition can be embodied into  the quadratic term $-r [p(t)]^2/K$, where $K$ is the carrying capacity. 
This negative feedback of the population on the growth rate
$r \to r (1-p(t))/K$ leads to a cross over from the exponential
growth for $p(t) \ll K$ to a saturation of the population at long times, which asymptotically converges to $K$.
Verhulst thought that Malthus was wrong (and therefore over-pessimistic when 
comparing human growth with food resources) not to take into account the negative feedbacks
embodied in the  quadratic term $-r [p(t)]^2/K$, that would lead naturally to an equilibrium.

But, the human population at the time of Verhulst  and until around 1960
has followed neither his specification, nor the Malthusian exponential growth.
As reviewed by \cite{Sornette2001} and references therein, the human population has grown faster than exponential, with 
the growth rate $r$ growing itself.

The simplest generalization of equation (\ref{eq:malthus}) that accounts for this observation 
assumes that the growth rate $r$ becomes $r \cdot [p/p_0]^\delta$, where $\delta >0$ and $p_0$ is
some reference population. The positivity of $\delta$ captures the positive feedback of population
on the growth rate: the larger the population, the larger the growth rate! Equation (\ref{eq:malthus}) then transforms into
\begin{equation}
\label{eq:delta}
\frac{d p}{d t} = R \cdot p(t)^{1+\delta} d t~,
\end{equation}
where $R=r/p_0^\delta$.
The solution of equation (\ref{eq:delta}) reads
\begin{equation}
\label{eq:power} 
p(t) \propto (t_c - t)^{-1/\delta}	 \quad \quad  \text{if } \delta > 0~.
\end{equation}
Here, the critical time $t_c$ at which the solution diverges is determined
from the parameters of equation (\ref{eq:delta}) and the initial population.
For $\delta=0$, we recover the exponential solution (\ref{eq:expsol}), seen as the 
limit of a finite-time-singularity (FTS) power law with exponent tending to zero.
The singular solution (\ref{eq:power}) was first discussed by von Foerster et al.
\cite{vonFoerster1960} (see \cite{Umpleby90} for assessments of the relative
merits of the ``natural science'' versus the ``demographic'' approach,
\cite{Kremer93} for an economic underpinning that we explore later, and 
\cite{Sornette2001,Korotayevetal06} for extensive generalizations).
In ecology, the positive correlation between population density and the per capita population growth rate
at the origin of the FTS behavior (\ref{eq:power})
is known as the Allee effect, see for instance \cite{Alleeeffect1999}. More generally, Allee discovered the existence of
an often present positive relationship between some component of individual fitness and either
numbers or density of conspecifics. The Allee effect is usually used to refer to 
the self-reinforcing feedbacks that promote accelerate extinction of species,
that can be modeled by finite-time crossing of zero, see \cite{Yukalov09}. %\cite{Yukalov10}
Goriely provides a rigorous mathematical framework \cite{Goriely2000} 
with a generalized version of equation~(\ref{eq:delta}), where 
the right hand side is replaced by an arbitrary polynomial of $p(t)$.

The use of the mathematics of FTS to describe and diagnose changes of regime is not new.
For instance, we refer to \cite{Sornette2001, Sornettecrash2003} for population dynamics and financial markets, 
\cite{SamSor02} for applications to engineering failures and earthquakes, \cite{SorprediPNAS02, Sornette2006Critical} for a large variety of systems,  \cite{Dakosetal08} for climate systems, %, \cite{ThompsonSieber10}
and
\cite{Scheffer2009, Biggsetal09, Drakeetal10} for environmental systems. These authors 
applied the concept of dynamical phase transitions and FTS to 
different systems exhibiting a bifurcation, crisis, catastrophe or tipping point, by showing how
specific signatures can be used for advance warnings.

One can generalize (\ref{eq:delta}) to take into account positive feedbacks
of the growth rate ${d \ln p}/{d t}$  on its rate of change ${d^2 \ln p}/{d t^2}$ (see \cite{IdeSor02}),
to arrive at solutions that exhibit FTS not in the variable $p(t)$, but in its derivative
${d p}/{d t}$. We will thus use the slightly more general expression
encompassing these cases:
\begin{equation}
p_{\rm power}(t) = a (t_c - t)^{-1/\delta} +c~.
\label{eq:deltafinite2}
\end{equation}
A FTS in ${d p}/{d t}$ and not in $p(t)$ corresponds to $-\infty < \delta < -1$ such that $0 < -1/\delta < 1$,
together with $a<0$ for an increase up to the value $p_{\rm power}(t_c) = c$. Here, the meaning of the
exponent $\delta$ is different from its use in equation (\ref{eq:delta}).

We shall use the  \emph{exponential model} (\ref{eq:expsol}) and the \emph{power law model} (\ref{eq:deltafinite2}),
as our two competing hypotheses. The essential difference between the \emph{exponential model} and the power law model is that the former
is defined for all times, while the later is valid only up to a finite time, the critical time $t_c$ beyond which
the solution ceases to exist. The singular behavior at $t_c$ is not meant to predict a genuine divergence
but only, as already stressed, that the system is exhibiting a transition to a qualitatively new regime.

%%%%%%%%%%%%%%%%%%%%%%%%%%%%%%%%%%%%%%%
\subsection{Properties distinguishing the exponential and the power law model}

Heated discussions among demographers greeted
the publication of the paper \cite{vonFoerster1960} concerning the singular solution (\ref{eq:power}):
the demographers
criticized the use of mathematical models such as (\ref{eq:delta}) as perhaps the clearest illustration of 
how bad use of mathematics may yield senseless results; actually, what the demographers missed
was that the FTS should not be taken at face value, but as the signature of a transition
to a new regime. Singularities do not exist in natural and social systems, but the singularities
of our approximate mathematical models are usually very good diagnostic of the change of 
regimes that occur in these systems.  The perhaps clearest examples are the phase transitions
between different states of matter (solid-liquid-gas-plasma, magnetized to non-magnetized, and so on)
that statistical physics describes so well with its classification involving the nature of the
singularity exhibited by the free energy of the system \cite{Goldenfeld92}.

As $t$ approaches $t_c$ from below,
%\footnote{We are only discussing the physically relevant case of $t \nearrow t_c$, the properties for $t \searrow t_c$ could be obtained analogously.},  
two regimes can be observed for the power law model:
\begin{itemize}
\item[$\delta<0$:] $(t_c-t)^{-1/\delta}$ goes to zero for $t \rightarrow t_c$ and $\pPower \rightarrow c$.
\item[$\delta>0$:] $(t_c-t)^{-1/\delta}$ goes to infinity for $t \rightarrow t_c$ and $\pPower \rightarrow \text{sign}(a) \cdot \infty$.
\end{itemize}

\begin{figure}
  \begin{center}
\includegraphics[angle=-90, width=8.7cm]{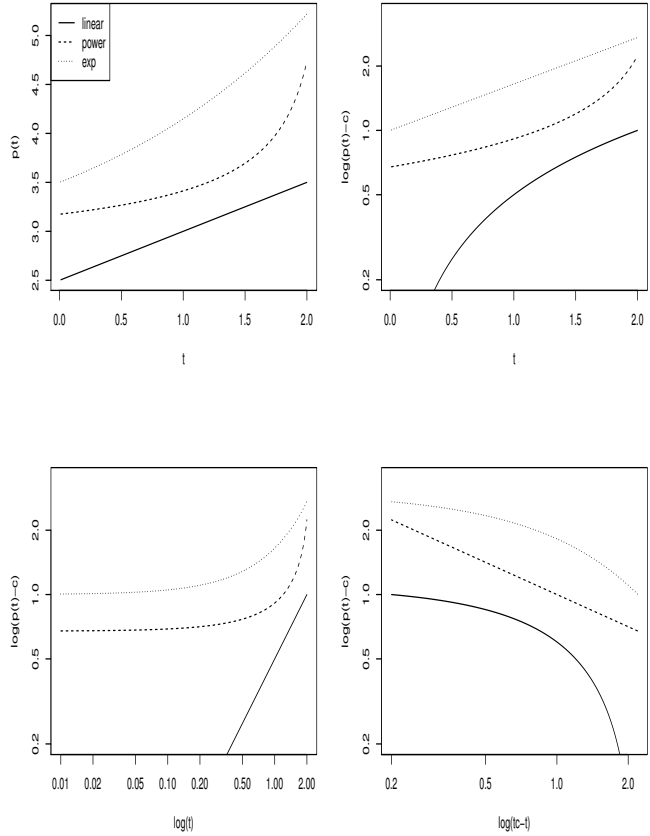}

\end{center}
\caption{Illustration of the qualitatively different behaviors of the \emph{exponential model} (\ref{eq:expsol}), the \emph{power law model} (\ref{eq:power}) and a \emph{linear model}, in different standard plot representations. For each of the four plots, the linear function $0.5 t+3.25$ is compared with the exponential function $1e^{0.5t}+2.5$ and with the power law $1(2.2-t)^{-0.5}+2.5$. (a)~is linear-linear, (b)~is linear-log, (c)~is log-log and (d)~is log-log referenced to the singularity. The constant $c$ is set equal 2.5. The relative vertical
positions of the three curves are arbitrarily chosen (from the above values) for the sake of a clear visualization.}
\label{fig:comparison}
\end{figure}

Figure~\ref{fig:comparison} in the SI illustrates the qualitatively different behaviors 
allowing one to distinguish between the linear growth model ($dp(t)/dt \sim t$), the
 \emph{exponential model} (\ref{eq:expsol}) and the \emph{power law model} (\ref{eq:deltafinite2}),
in different standard plot representations. 

%%%%%%%%%%%%%%%%%%%%%%%%%%%%%%%%%%%%%%%
\subsection{Faster-than-exponential growth by feedbacks between macro-economic variables}

Up to now, we have postulated the form (\ref{eq:delta}) to capture the possible existence of a positive feedback of population
on the population growth rate. Such a simplified ansatz leaves two issues unresolved. First, the positive feedback
of population on growth rate may not be direct, but mediated by other variables via indirect mechanisms. 
Second, the consequences on the dynamics of carbon dioxide emissions are not clear. We thus address these two issues
using an economic framework developed by Kremer \cite{Kremer93}, following the approach of Johansen and Sornette \cite{Sornette2001}.
The following derivation is not intended to represent a faithful economic growth model that
we would like to promote, but is offered
to illustrate the importance of indirect mechanisms in growth processes. In particular, 
we would like to stress the fact
that faster-than-exponential growth is a robust outcome of multi-dimensional loop processes.
Even when each feedback process individually leads to an exponential or even a subdued sub-exponential
growth, the overall  dynamics can be super-exponential.

In economics, population $p(t)$ translates into labor force $L(t)$, which is assumed to be proportional to $p(t)$. 
In addition to population represented by the labor force, we consider the effect of technology level $A(t)$
and of the amount $K(t)$ of available capital. In the presence of labor and capital, with a given technology level, 
the economy is going to produce an output $Y(t)$, for instance proxied by GDP.
In the macroeconomics of endogenous growth \cite{Romer2000}, 
it is common to use the Cobb-Douglas equation (originally developed by \cite{Cobb1928} and extensively discussed in \cite{Romer2000}) to relate the total output to labor, capital and technology
as follows\footnote{$A'$ in the IPAT equations stands for gross world product per capita, whereas in the Cobb-Douglas equation $A$ stands for technology. Further, the IPAT equation uses $T$ instead of $A$ to denote technology. Similar, the macro-economists refer to $L$ as labor, whereas $P$ in the IPAT equality stands for population. We will not distinguish between labor $L$ and population $P$ and use the terms interchangeably.}:
\begin{equation}
\label{eq:CobbDouglas}
Y(t) =  K(t)^\alpha (A(t) L(t))^{1-\alpha}~, \textrm{with~}~0 < \alpha <1~.
\end{equation}
Furthermore, we use the assumption by Solow that a constant fraction $s$ of the economy goes to savings ,
i.e.~capital grows according to 
\begin{equation}
\label{eq:AssumptionSolow}
\frac{dK}{dt} = s Y(t)~.
\end{equation}
Following \cite{Kremer93}, we assume that, as already mentioned, labor is proportional to capital 
\begin{equation}
\label{eq:AssumptionKremer}
K(t) \sim L(t)~.
\end{equation}
We further assume that technology change is depending on capital, labor and current level of technology
according to
\begin{equation}
\label{eq:AssumptionTechnology}
\frac{dA}{dt} = d K(t)^\eta \times L(t)^\gamma \times A(t)^\theta~,
\end{equation}
where the exponents $\eta, \gamma$ and $\theta$ are all positive, expressing a positive feedback
effect of each of the variables on the growth of technology.
Putting together all these ingredients, we can rewrite the 
 Kremer (\ref{eq:AssumptionKremer}) and Solow (\ref{eq:AssumptionSolow}) equations as
 a system of two coupled ordinary differential equations: 
\begin{align}
\frac{dA}{dt} & = e L(t)^{\eta + \gamma} \times A(t)^\theta, \label{eq:ODE_system1} \\
\frac{dL}{dt} & = f L(t) \times A(t)^{1-\alpha}~. \label{eq:ODE_system2}
\end{align}
Equation (\ref{eq:ODE_system2}) basically states that labor (and thus population) 
is growing exponentially, holding technology constant. In other words, the growth rate
of population is controlled by a nonlinear function of technology. Here, this nonlinear function
is a power law with exponent $0 < 1-\alpha <1$, which embodies the benefits that
technology brings in decreasing death rates, for instance via improvement
in health care. Invoking this mechanism is standard in demographic research.

We look for solutions exhibiting a FTS of the form 
\begin{align}
\label{eq:solutionTechnology}
A(t) & = A_0 (t_c - t)^{-1/\mu}~,\\
\label{eq:solutionLabour}
L(t) & = L_0 (t_c - t)^{-1/\kappa}~.
\end{align}
Note that the critical time $t_c$ of the singularity, if it exists, is necessarily the same for both variables, 
as seen from inspection of the two coupled  equations (\ref{eq:ODE_system1},\ref{eq:ODE_system2}).
Inserting this ansatz in equations (\ref{eq:ODE_system1},\ref{eq:ODE_system2}),
%\footnote{The exact derivation is given in the SI.}
we obtain
a system of differential equations for the unknown exponents $\mu$ and $\kappa$, whose solutions read
\begin{align}
\mu & = 1-\alpha~,  \label{eq:SolutionDelta}\\
\kappa & = \frac{\eta +\gamma}{2-\theta-\alpha} (1-\alpha)~.
\label{eq:SolutionKappa}
\end{align}
The condition for the solutions~(\ref{eq:solutionTechnology},\ref{eq:solutionLabour}) to hold
is that $\mu$ and $\kappa$ be strictly positive. This implies $0 < \alpha <1$ and $\alpha < 2 -\theta$.
If $\theta  \leq 1$, then the conditions are always satisfied in the regime where the 
Cobb-Douglas equation holds.  The case $\theta \leq 1$ is particularly interesting because
it corresponds to a sub-exponential growth of technology, for a fixed labor force. 
In other words, for a fixed population level, equation (\ref{eq:ODE_system1}) gives
a long-time growth of the form $A(t) \sim t^{\frac{1}{1-\theta}}$, which is sub-exponential
(slower than exponential) for $\theta < 1$ and exactly exponential for $\theta=1$.
It is the coupling between a sub-exponential growth of $A(t)$
and an exponential growth of population $L(t)$ mediated by nonlinear feedback  loops
that create the super-exponential finite-time singularity. This behavior underlies the possible
traps of single variable analysis.

These results can be translated into a prediction of carbon dioxide emission via the following simple assumption.
Assuming that carbon dioxide emissions
are proportional to production divided by some power of technology $\xi$, we have
\begin{equation}
\label{eq:solution}
\frac{d\text{CO}_2}{dt} = \frac{Y(t)}{A(t)^\xi} = h  (t_c-t)^{-1/\varphi}, 
\end{equation}
where $\varphi = (1/\kappa - \xi/\mu + 1)^{-1}$ (see SI for details of the derivation)
and $\text{CO}_2$ stands for the total carbon dioxyde content in the atmosphere.
The introduction of a non-zero exponent $\xi$ accounts
for the common observation that more developed countries tend to have a lower footprint and smaller
carbon emissions per unit of output, due to the progressive adoption of more efficient technologies and 
the increasing importance of a clean environment in the utility functions of consumers.
%and as long as the combination. of exponents ensures that ???? $(\frac{\eta +\gamma}{2-\theta-\alpha} - \xi) (1-\alpha)  + 1 >0$.

Let us thus stress the main result of this exercise.  We have  $\frac{dA}{dt} \sim  A(t)^\theta$
at fixed labor with $\theta <1$ and $\frac{dL}{dt} \sim  L(t)$ at fixed technology. Thus, there is no way to get
a faster-than-exponential growth in any of these two variables alone. However, when coupling them
via the feedback of labor on technology and that of technology on labor, the
FTS power law solutions (\ref{eq:solutionTechnology},\ref{eq:solutionLabour}) emerge.
Hence, a finite-time singularity can be created from the interplay of several growing variables resulting in a non-trivial behavior: the interplay between different quantities may produce an ``explosion'' in the population even though the individual dynamics do not! 

Of course, infinities do not exist on a finite Earth!  
These singularities should not be interpreted as the prediction of real ``blow-ups''. They can be however
faithful description of the transient dynamics up to a neighborhood of the predicted critical time $t_c$.
Around $t_c$, new mechanisms kick in and produce a \emph{change of regime}.

\begin{figure}
\begin{center}
\includegraphics{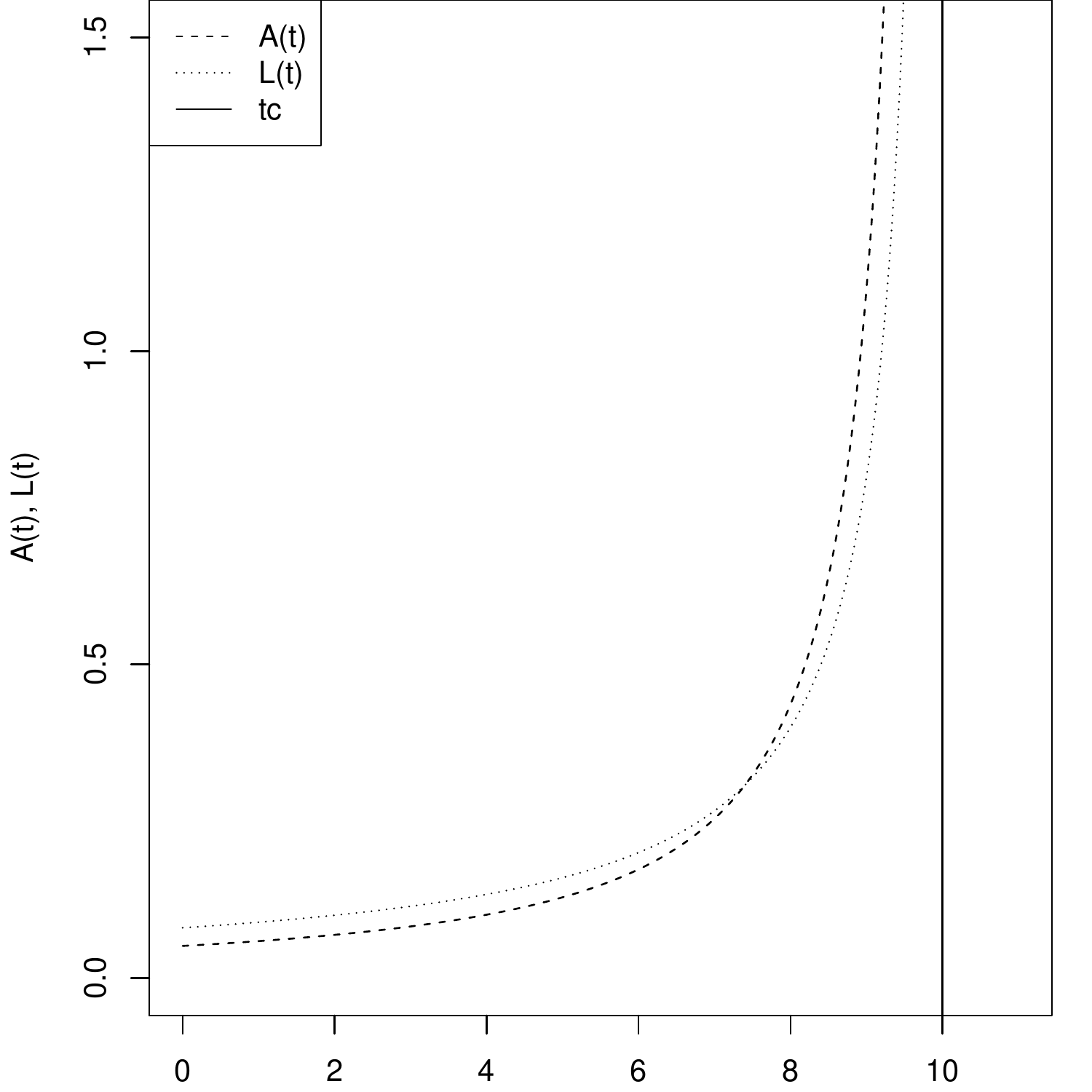}
\end{center}
\caption{ 
Numerical solution of equations~(\ref{eq:ODE_system1}) and (\ref{eq:ODE_system2}) with $\alpha=\frac{1}{4}$, $\eta + \gamma=1$, $\theta=1$ and $t_c=10$. The  initial conditions are $A(0)=1.1$ and $L(0)=0.8$.
We assume without loss of generality $e=f=1$, as these coefficients can be absorbed in the units of $A$ and $L$ respectively.
$L(t)$ and $A(t)$ grow super-exponentially towards a singularity occurring at the same
time as a result of their coupling. The logarithms of $A(t)$ and $L(t)$ are plotted as a function of (linear time). 
The upward curvatures and approaches to the singular vertical asymptote exemplify the super-exponential growth.
}
\label{fig:ODE_system}
\end{figure}

To illustrate the above point, let us go through a detailed scenario where the individual processes stay finite in finite time, but the combination via feedback can lead to finite time singularities.
Consider the following parameters
\begin{description}
  \item[$\alpha = \frac{1}{4}$:] as in the seminal paper \cite{Cobb1928}.
  \item[$\theta = 1$:] Linear feedback from technology $A$ on itself. Holding all other factors constant, technology will grow exponentially (see equation (\ref{eq:AssumptionTechnology})).
  \item[$\eta + \gamma = 1$:] The simplest possible, non-trivial, assumption. 
\end{description}
With these numbers, we obtain the two exponents  $\mu = 3/4$ and $\kappa = 1$ for the equations (\ref{eq:solutionTechnology}) and (\ref{eq:solutionLabour}), respectively, and the value $1/\varphi = 5/3$ for the rate of carbon dioxide emission given by 
equation (\ref{eq:solution}), assuming carbon dioxide emission per capita technology is as efficient as general technology $A$, i.e.~$\alpha=\xi=1/4$. Although, we have only assumed exponential growth of all individual factors, carbon dioxide emission is 
predicted in this example to grow faster than exponential, leading to a mathematical FTS which is the 
signature of a non-sustainable regime towards a new behavior (see Figure~\ref{fig:ODE_system}).

Even less stringent conditions for a FTS to occur are 
needed when the description of the dynamics of the system in terms of two coupled  equations (\ref{eq:ODE_system1},\ref{eq:ODE_system2}) is augmented to take into account the dynamics of additional coupled variables, leading
to systems of three or four coupled equations. Such additional positive
feedback loops include nonlinear lagged dependencies of capital on labor (thus extending Kremer's simplifying assumption 
(\ref{eq:AssumptionKremer})).

%%%%%%%%%%%%%%%%%%%%%%%%%%%%%%%%%%%%%%%%%%%%%%%%%%%%%%%%%%%%%%%%%%%%%%%%%%%%%%%
\section{Empirical tests on human population and atmospheric carbon dioxide content}
%%%%%%%%%%%%%%%%%%%%%%%%%%%%%%%%%%%%%%%
\subsection{Population}
\begin{figure}
\begin{center}
\includegraphics{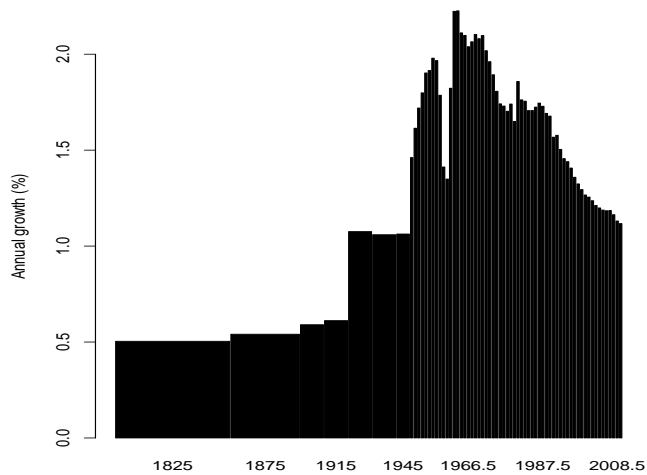}
\end{center}
  \caption{Annualized world population growth rate from year 1800 -- 2010.}
  \label{fig:population_percent}
\end{figure}

Figure~\ref{fig:population_percent} shows that the growth rate of the World population was a strongly
increasing function of time till the late 1950s.
A sharp decrease of the growth rate occurred, then followed by a resumed acceleration
till its peak in 1964, from which a slow decrease can be observed.

\begin{figure}
\begin{center}
\includegraphics{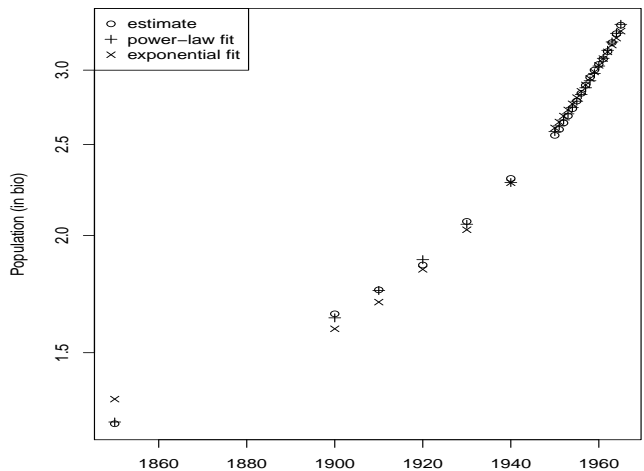}
\end{center}
\caption{Population data represented by the empty circles (where ``estimate'' refers
to the empirical estimation of the population) 
fitted over the time window from 1850 -- 1965  by the FTS power-law (\ref{eq:deltafinite2}) and the exponential model (\ref{eq:expsol}).
The fitted parameters are $\delta=2$ and  $t_c=1988$ 
for the power-law and $r=0.028$ for the exponential fit.% (corresponding to a doubling time of about $35$ years according to the exponential fit).
%The ratio of squared errors between the power-law and the exponential-fit is $0.18$.
}
\label{fig:population_fit_early}
\end{figure}

The first regime till about 1960 is incompatible with the exponential model, which corresponds
to a constant growth rate. Figure~\ref{fig:population_fit_early} shows that, over the time period 1850 to 1965,
the exponential model is inferior to the FTS power law model.
Using model (\ref{eq:delta}), we estimate that the growth exponent 
$\delta$ is approximately equal to 2, that is, even larger than the value $1$ estimated by 
\cite{vonFoerster1960}: clearly, population growth over this time period was faster
than exponential and the FTS power law model accounts parsimoniously for the data.

%%+++++++++++++++++++++++++++++++++++++++++++++++++++++++++++++++++++++++++++++
% population_fit_early
%%+++++++++++++++++++++++++++++++++++++++++++++++++++++++++++++++++++++++++++++

\begin{figure}
\begin{center}
\includegraphics{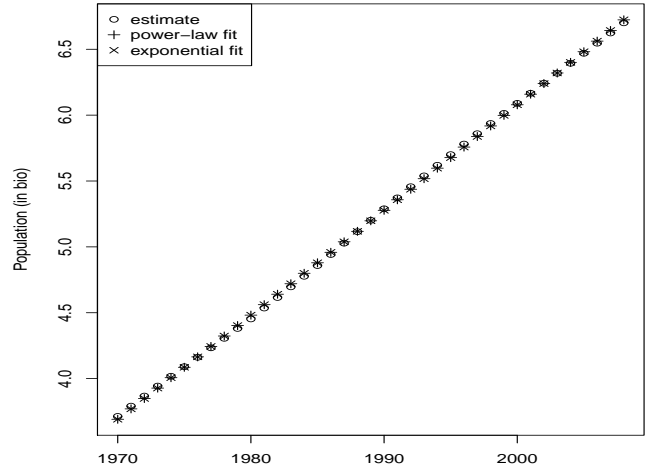}
\end{center}
\caption{Population data fitted over the time window from 1970 -- 2008 by the FTS power-law (\ref{eq:deltafinite2}) and the exponential model (\ref{eq:expsol}).
The fitted parameters are $\delta=3.5$ and  $t_c=3939$ 
for the power-law and $r=0.00067$ for the exponential fit. %  (corresponding to a doubling time of about $80$ years according to the exponential fit.
%The ratio of squared errors between the power-law and the exponential-fit is $1$.
%VERIFIE THE VALUE OF $r$
}
\label{fig:population_late}
\end{figure}

Figure~\ref{fig:population_late} shows that, over the time period from 1970 to 2008, the exponential model (\ref{eq:expsol})  and the FTS power law model (\ref{eq:deltafinite2}) are indistinguishable. 
By Occam's razor, the exponential model with an approximately constant growth rate is then preferred.

%%+++++++++++++++++++++++++++++++++++++++++++++++++++++++++++++++++++++++++++++
% population_late
%%+++++++++++++++++++++++++++++++++++++++++++++++++++++++++++++++++++++++++++++

%%%%%%%%%%%%%%%%%%%%%%%%%%%%%%%%%%%%%%%
\subsection{Carbon Dioxide content in the atmosphere}

%Detailed data on the quantity of carbon dioxide in the atmosphere is available since 1959, when air measurements began at Mauna Loa, in Hawaii archipelo.
%This data was obtained from the web site of the Earth System Research Laboratory\footnote{\url{http://www.esrl.noaa.gov/gmd/ccgg/trends/}}.
%Longer term data is only available from indirect measurements, i.e.~ice core data.
%The data for long time periods was retrieved from the web site of the Carbon Dioxide Information Analysis Center web site\footnote{\url{http://cdiac.ornl.gov/}}\footnote{and some other sources\ldots}.

\begin{figure}
\begin{center}
\includegraphics{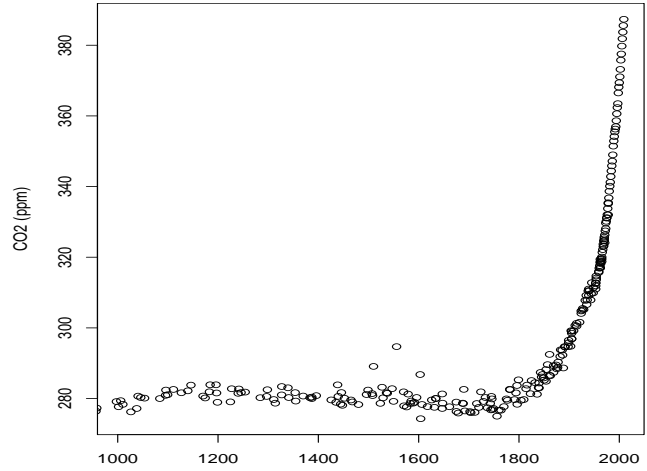}
\end{center}
\caption{Atmospheric carbon dioxide since 1000 CE to present. The data shown combines ice core and air measurements from different sources. See data section for more details.}
\label{fig:co2_overview}
\end{figure}

Figure~\ref{fig:co2_overview} in the SI plots the carbon dioxide content in 
the atmosphere since 1000 CE\@. The dramatic acceleration due to 
anthropogenic forcing since the 1800s is clearly observed.

\begin{figure}
\begin{center}
\includegraphics{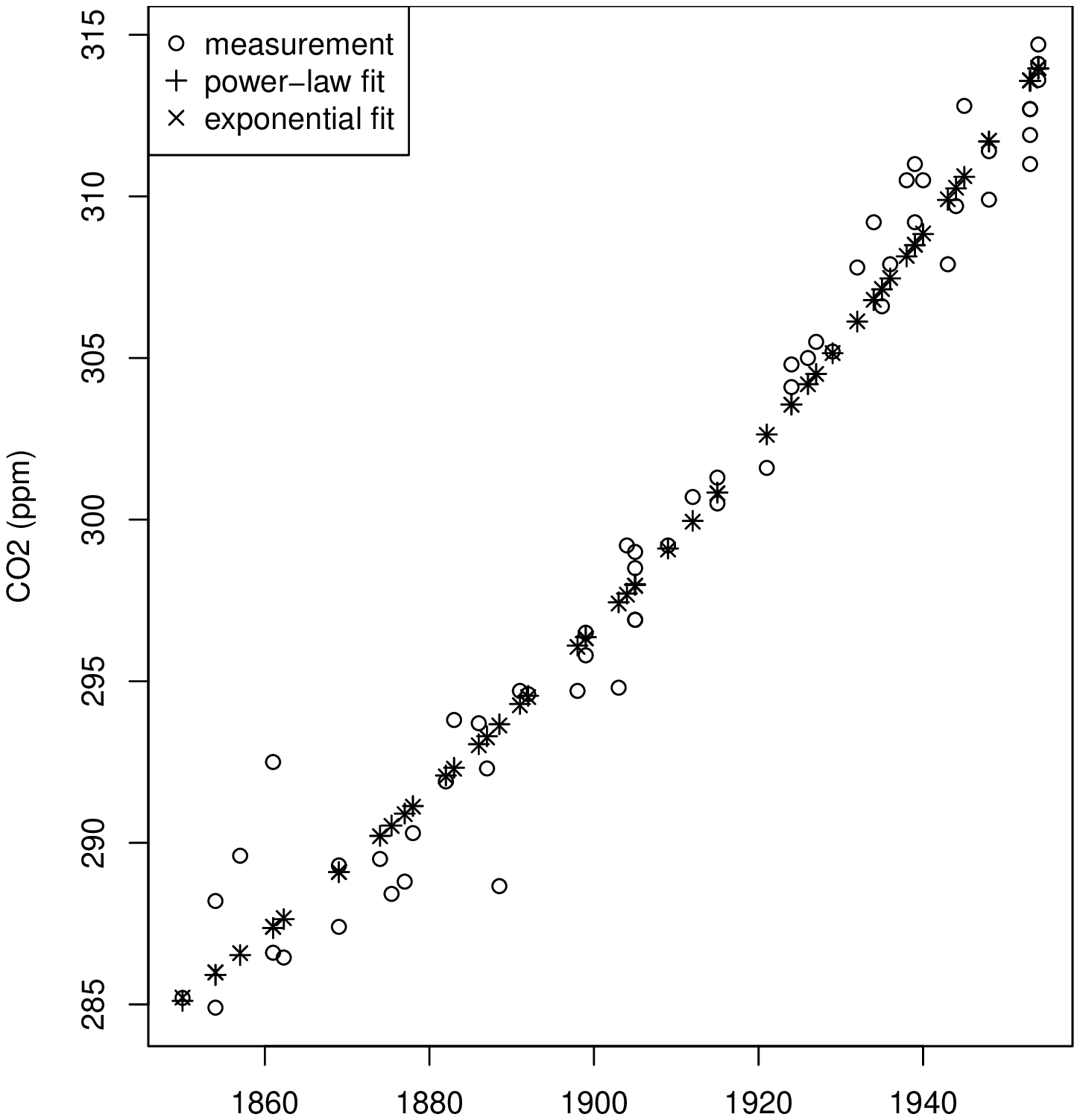}
\end{center}
\caption{Carbon dioxide data fitted over the time window from 1850 -- 1954 by the FTS power-law (\ref{eq:deltafinite2}) and the exponential model (\ref{eq:expsol}).
The fitted parameters are $\delta=0.65$ and $t_c=2304$ 
for the power-law and $r=0.0066$ for the exponential fit.
%(corresponding to a doubling time of about $281$ years according to the exponential fit).
The two fits are almost undistinguishable and their goodness-of-fit is essentially the same. 
%The ratio of squared errors between the power-law and the exponential-fit is $1$.
}
\label{fig:co2_fit_early}
\end{figure}

\begin{figure}
\begin{center}
\includegraphics{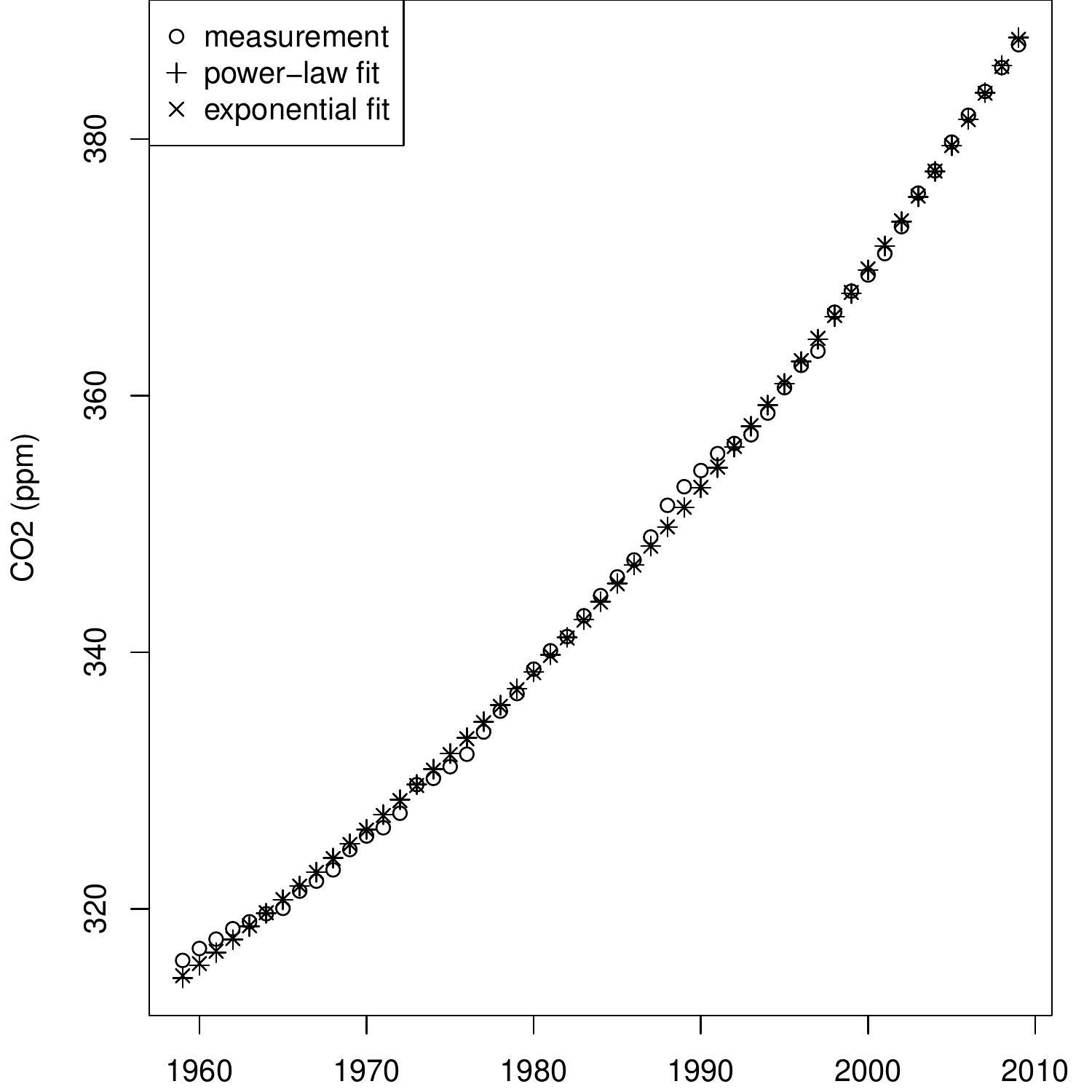}
\end{center}
\caption{Carbon dioxide data fitted over the time window from 1959 -- 2009 by the FTS power-law (\ref{eq:deltafinite2}) and the exponential model (\ref{eq:expsol}).
The fitted parameters are $\delta=0.73$ and  $t_c=2132$ 
for the power-law and $r=0.016$ for the exponential fit. %(corresponding to a doubling time of about $84$ years according to the exponential fit).
The two fits are almost undistinguishable and their goodness-of-fit is essentially the same. 
%The ratio of squared errors between the power-law and the exponential-fit is $1.1$.
}
\label{fig:co2_fit_late}
\end{figure}

We calibrate the \emph{exponential model} (\ref{eq:expsol}) and the \emph{power law model} (\ref{eq:deltafinite2})
separately to two time periods: (i) from 1850
to 1954 (Figure~\ref{fig:co2_fit_early}), for which the data originates from ice drill cores and (ii)
from 1959 to 2009 (Figure~\ref{fig:co2_fit_late}), for which the data originates from air
samples. %\footnote{We leave out the somehow turbulent years of 1941 -- 1954, which are distorted due to WWII and its aftermath.}
The quality of the fits by the two models, as quantified by the 
sum of squared errors between theory and data, is practically equivalent.
Therefore, we cannot reject the hypothesis that the exponential model is sufficient
to explain the data for each time window \emph{separately}. 

\begin{figure}
\begin{center}
\includegraphics{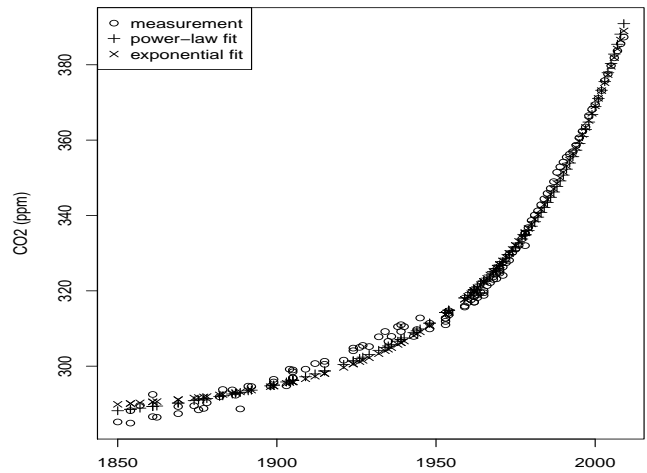}
\end{center}
\caption{Carbon dioxide data fitted over the time window from 1850 -- 2009 by the FTS power-law (\ref{eq:deltafinite2}) and the exponential model (\ref{eq:expsol}).
The fitted parameters are $\delta=0.33$ and $t_c=2129$ 
for the power-law and $r=0.024$ for the exponential fit. %(corresponding to a doubling time of about $65$ years according to the exponential fit).
The ratio of squared errors between the power-law and the exponential-fit is $0.88$.
}
\label{fig:co2_all}
\end{figure}

However,
the growth rate $r$ calibrated with the exponential model (see equation~(\ref{eq:expsol}) has more than doubled
from the first period 1850 -- 1954 ($r=0.0066$) to the second period
1959 -- 2009 ($r=0.016$). While being not fully warranted given the heterogeneity of the data sources,
we have fitted the two models to the whole period from 1850 to 2009. 
We find that the FTS power law is the clear winner (see Figure~\ref{fig:co2_all}) which, together
with the more than doubling of the growth rate $r$ from the first to the second time intervals, suggests
the existence indeed of a faster-than-exponential growth of the atmospheric content of carbon dioxide.

%%+++++++++++++++++++++++++++++++++++++++++++++++++++++++++++++++++++++++++++++
% co2_fit_early
%%+++++++++++++++++++++++++++++++++++++++++++++++++++++++++++++++++++++++++++++

%%+++++++++++++++++++++++++++++++++++++++++++++++++++++++++++++++++++++++++++++
% co2_fit_late
%%+++++++++++++++++++++++++++++++++++++++++++++++++++++++++++++++++++++++++++++

\begin{figure}
\begin{center}
\includegraphics{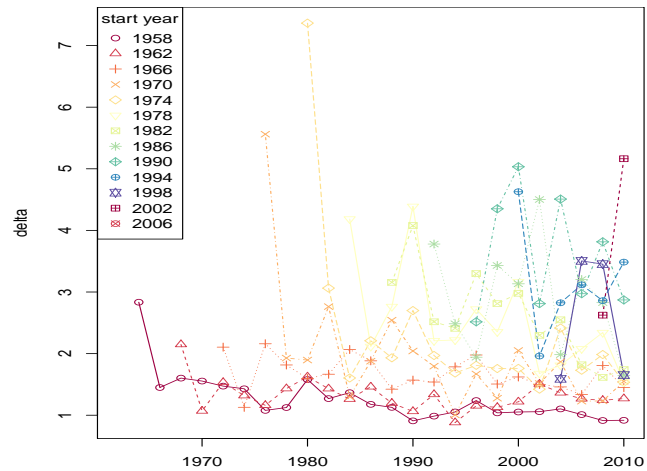}
\end{center}
\caption{Estimates of the exponent $\delta$ of equation (\ref{eq:deltafinite2}) on the monthly Mauna Loa carbon dioxide data obtained from air measurements in different intervals $[t_1, t_2]$. 
Each line corresponds to a specific start time $t_1$, as shown in the legend. The ending point $t_2$ is the variable on the abscissa.}
\label{fig:co2_special1}
\end{figure}

We now attempt to be more precise on the nature and evolution of the faster-than-exponential growth by
estimating the exponent $\delta$ of equations (\ref{eq:deltafinite2}) applied to the time series of carbon dioxide
atmospheric content. We use the monthly data from the Mauna Loa site, as it is considered to be one of the most reliable.
Before calibrating equation (\ref{eq:deltafinite2}) to various time intervals $[t_1, t_2]$, we smooth the data
by using a Gaussian kernel with a width of $10$ years. Then, we estimate $\delta$,
with $t_1$ being scanned from 1958 to 2006 and $t_2$ 
being scanned from 1960 to 2009 as shown in Figure \ref{fig:co2_special1}.

Two main results are obtained. First, the exponent $\delta$ is found almost always larger than or equal to $1$,
implying a growth at least as fast as exponential and often significantly faster. 
Second, one can observe a systematic trend. For time intervals
starting earlier (i.e., for $t_1$ in the late 1950s and in the 1960s), the exponent $\delta$ tends to be closer to $1$, while 
for larger $t_1$, $\delta$ is significantly larger than $1$. This leads to the conclusion that
the carbon dioxide content in the Earth atmosphere is growing at least exponential and probably
faster-than-exponentially, with no sign of abating. The latest time intervals are characterized
by the largest exponents $\delta$'s, significantly above the lower bound $0$ that would correspond
to an exponential growth. The content of carbon dioxide
in the atmosphere is accelerating super-exponentially.

%+++++++++++++++++++++++++++++++++++++++++++++++++++++++++++++++++++++++++++++
% co2_special1
%+++++++++++++++++++++++++++++++++++++++++++++++++++++++++++++++++++++++++++++

%%%%%%%%%%%%%%%%%%%%%%%%%%%%%%%%%%%%%%%
\subsection{Compatibility between exponential population growth and super-exponential CO2 emissions}

The previous empirical evidence suggests that the human population on the Earth is growing now just
exponentially, while there is suggestive evidence that the content of carbon dioxide
in the atmosphere is accelerating super-exponentially. How are these two different behaviors be compatible with 
the solutions (\ref{eq:solutionTechnology},\ref{eq:solutionLabour}) for $A(t)$ and $L(t)$ of equations
(\ref{eq:ODE_system1},\ref{eq:ODE_system2})? 

We consider two possible explanations. The first one would argue that until the 1960s both
population and atmospheric carbon dioxide content were super-exponentially accelerating
in accordance with expressions (\ref{eq:solutionTechnology},\ref{eq:solutionLabour}). 
Then, the slowing down from super-exponential to just exponential growth of the human population
could be interpreted as a finite-size effect that is starting to be felt for this variable only, as physical limits
are more stringent for the human carrying capacity and the response of human birth and death
rates to policies than they are for carbon dioxide emissions.

The second explanation is that the two different behaviors of $A(t)$ and $L(t)$ may be resolved
within the mathematical structure developed in equations~(\ref{eq:solutionTechnology}) and~(\ref{eq:solutionLabour}). Indeed, let us assume that 
the growth of the human population is following solution (\ref{eq:solutionLabour}), but with a small
value of the exponent $\kappa$. For all practical purpose, a FTS power law with 
a small exponent is indistinguishable from
an exponential growth over a finite time interval. This interpretation is reasonable in so far
that human population growth has been unambiguously super-exponential until the 1960s, and it is only
recently that this growth has abated. %It is thus quite possible that it is still super-exponential
%but to a degree that is not sufficiently strong to be distinguishable from a pure exponential, as
%shown in the analysis of Figures~\ref{fig:population_fit_early} and~\ref{fig:population_late}. 

Let us now turn to the dynamics of $\text{CO}_2$ content. The conditions for a super-exponential growth of the 
content of carbon dioxide in the atmosphere are compounded by many complex processes
involving, in addition to the emissions, the sequestrations of $\text{CO}_2$ by, and dynamics of, the ocean and biosphere.
As a rough rule of thumb, 
we assume that the  total content of carbon dioxide in the atmosphere 
at time $t$ is simply proportional to (but likely less than)
the cumulative release of $\text{CO}_2$ until time $t$. In other words, $\text{CO}_2$ content is estimated as
a finite fraction of the solution of equation (\ref{eq:solution}). Under these assumptions,
in order for $\text{CO}_2$ content to exhibit a FTS power law behavior,
it is necessary and sufficient that the exponent $1/\varphi$ in (\ref{eq:solution}) be larger than $1$.
Indeed, by integration, $\text{CO}_2 (t)$ remains of the same form $(t_c-t)^{-1/\delta}$, with 
$1/\delta=1/\varphi -1 >0$, where $\delta$ is defined as in equation~(\ref{eq:deltafinite2}). 
This condition translates into the condition $\xi < \mu / \kappa$. 
As we have assumed that $\kappa$ is small, corresponding to the closeness of the population
dynamics to an exponential growth, this condition does not provide a strong constraint for $\xi$:
$\text{CO}_2$ content can exhibit an (accelerated) FTS dynamics
even if $\xi$ is large, corresponding to a more efficient economy.
If $1-\alpha$ is close to $0$, corresponding to output mainly controlled by availability of capital,
then $\xi$ should be small. 
Small values of $\xi$ correspond to the situation in which, taken globally over the whole Earth,
the technological advances have not yet significantly abated carbon emission per unit of output.
This statement may appear shocking and counter-factual for developed countries.
But, at the scale of the whole planet, one can observe that improvement in carbon emissions
(i.e., decrease per unit of output)
in the developed countries are counteracted by the increases of carbon emissions
in some major developing countries, such as China, India and Brazil, which use carbon emission inefficient
technologies (for instance heavily based on coal burning).
In summary, we find a very robust FTS behavior for $\text{CO}_2$ over a broad and
realistic range of parameters, which makes it difficult to constrain the impact of the advance
of technology on production efficiency.

%%%%%%%%%%%%%%%%%%%%%%%%%%%%%%%%%%%%%%%%%%%%%%%%%%%%%%%%%%%%%%%%%%%%%%%%%%%%%%%
\section{Conclusion}

We have analyzed the growth of atmospheric carbon dioxide and of 
what constitutes arguably its most important underlying driving variable, namely human population. 
Our empirical calibrations suggest that human population has decelerated
from its previous super-exponential growth until 1960 to 
``just'' an exponential growth. As for atmospheric $\text{CO}_2$ content, we find that it is at 
least exponentially 
increasing and more probably exhibiting an accelerating growth rate, consistent with
a FTS (finite-time singular) power law regime. 

We have proposed a simple framework to think about these dynamics, based on endogenous economic growth theory.
We showed that the positive feedback loops between several variables, such as population, technology and capital
can give rise to the observed FTS behavior, notwithstanding the fact that the dynamics of each variable
would be stable or at most exponential, conditional on the stationarity of the other variables. It is the
joint growth of the coupled variables that may give rise to the enormous acceleration characterized
by the FTS behavior, both in the equation and, we present suggestive evidence, in the carbon dioxide content in the atmosphere. 

Overall, the evidence presented here does not augur well for the future. 
\begin{itemize}
\item The human population is still
growing at a non-sustainable rate and there is no sign the population will stabilize anytime soon.
Many argue that economic developments and education of women will lead to a decrease growth rate
and an eventual stabilization of human population. This is not yet observed in the population dynamics,
when integrated worldwide. Let us hope that the stabilization of the human population will occur
endogenously by self-regulation, rather than by more stringent finite carrying capacity constraints
that can be expected to lead to severe strains on a significant fraction of the population.
\item Notwithstanding a lot of discussions, international meetings, prevalence in the media,
atmospheric $\text{CO}_2$ content growth continues unabated with a clear faster-than-exponential behavior.
On the face of this evidence using data until 2009, stabilizing atmospheric carbon dioxide emissions 
at levels reached in 1990 for instance
seems very ambitious, if not utterly unrealistic.
%\footnote{As another example, the initiative ``350'' (\url{http://www.350.org/mission}) sets the goal of bringing down atmospheric carbon dioxide to a level of 350ppm.}.
We are not pessimistic. We think that only evidence-based
decision making can lead to progress. The present evidence gives some measure of the 
enormous challenges to control our $\text{CO}_2$ emissions to acceptable levels.
\end{itemize}

\section{Data}

Population data was obtained from the website of the United Nations (\url{http://www.un.org/esa/population/publications/sixbillion/sixbilpart1.pdf}) and the website of U.S.~Census Bureau (\url{http://www.census.gov/ipc/www/idb/worldpop.php}). %see also overview at \url{http://www.census.gov/ipc/www/worldhis.html} % US Census bureau after 1950

Carbon dioxide data was collected from different sources: from the Carbon Dioxide Information Analysis Center (CDIAC) (\url{http://cdiac.esd.ornl.gov/ftp/trends/co2/siple2.013}, \url{http://cdiac.ornl.gov/ftp/trends/co2/lawdome.combined.dat}), the National Oceanic and Atmospheric Administration (NOAA) (\url{ftp://ftp.cmdl.noaa.gov/ccg/co2/trends/co2_annmean_mlo.txt}, \url{ftp://ftp.ncdc.noaa.gov/pub/data/paleo/icecore/antarctica/maud/edml-co2-2005.txt}, \url{ftp://ftp.cmdl.noaa.gov/ccg/co2/trends/co2_mm_mlo.txt}) and from \cite{Barnola1995}\footnote{leaving out ``Bern'' measurements}.

\begin{quote}
 We are grateful for stimulating discussion with Nicolas Gruber.  We acknowledge financial support from the ETH Competence Center ``Coping with Crises in Complex Socio-Economic Systems'' (CCSS) through ETH Research Grant CH1-01-08-2 and ETH Zurich Foundation.
\end{quote}

%%%%%%%%%%%%%%%%%%%%%%%%%%%%%%%%%%%%%%%%%%%%%%%%%%%%%%%%%%%%%%%%%%%%%%%%%%%%%%%
%\bibliographystyle{abbrvnat}
\bibliographystyle{plain}
\bibliography{ref}
%%%%%%%%%%%%%%%%%%%%%%%%%%%%%%%%%%%%%%%%%%%%%%%%%%%%%%%%%%%%%%%%%%%%%%%%%%%%%%%

\newpage

%\appendix[Supporting Information]
\appendix

%\renewcommand{\theequation}{SI \arabic{equation}}
%\setcounter{equation}{0}  % reset counter 

%%%%%%%%%%%%%%%%%%%%%%%%%%%%%%%%%%%%%%%%%%%%%%%%%%%%%%%%%%%%%%%%%%%%%%%%%%%%%%%
\section{Discussion of exponential growth / FTS power-law}

Depending on the scale of the abscissa and the ordinate, exponential growth and FTS power-law growth can look very different (see also Figure \ref{fig:comparison}):

\begin{itemize}
  
\item (a): the linear-linear plot shows the dual property of the FTS power law function, which is
to both grow initially slower than the two other models, and then to catch up explosively.

\item (b): in this linear-log plot, by construction, the exponential function is a straight-line, thus a linear dependence
in this representation qualifies an exponential growth. The linear model is concave (slower growth) and the 
power law FTS model is convex (faster growth).  

\item (c): the log-log plot would qualify a power law $t^\beta$ as a
straight line whose slope is the exponent $\beta$. Hence the linear function is also
linear in this representation with slope $1$. Both the exponential and FTS power law model
exhibit an upward convex shape. It is important not to confuse a power law and 
a FTS power law: the former is proportional to a power of $t$ 
and thus exists for all times, while the later is proportional
to a power of $t_c-t$ and is only defined for $t<t_c$. 

\item (d): in this log-log plot 
in the variable $t_c-t$, by construction, the FTS power law is qualified by a straight line behavior,
with a slope equal to the exponent $-1/\delta$. Both linear and exponential models
are associated with concave curves, characterizing a slower growth 
in the vicinity of $t_c$. Note that time $t$ increases to the left.
\end{itemize}

%%+++++++++++++++++++++++++++++++++++++++++++++++++++++++++++++++++++++++++++++
% comparison
%%+++++++++++++++++++++++++++++++++++++++++++++++++++++++++++++++++++++++++++++

%%%%%%%%%%%%%%%%%%%%%%%%%%%%%%%%%%%%%%%
\section{Exact Solution of the ODE system}

This appendix provides the exact derivation of the system of equations~(\ref{eq:ODE_system1}) and~(\ref{eq:ODE_system2}),
thus justifying the ansatz (\ref{eq:solutionTechnology}) and (\ref{eq:solutionLabour}) used.

First, we combine equations~(\ref{eq:ODE_system1}) and~(\ref{eq:ODE_system2}) into a single equation:
\begin{equation}
\frac{dA}{dt}  L(t)^{-\eta -\gamma} A(t)^{-\theta} - \frac{dL}{dt}  L(t)^{-1}  A(t)^{-1+\alpha} = 0~.
\end{equation}
Without loss of generality, we can set $e=f=1$ by defining appropriately the units of $A$ and $L$. 
Separating the variables and integrating lead to
\begin{equation}
\frac{1}{2-\alpha-\theta} A(t)^{2-\alpha-\theta} - \frac{1}{\eta+\gamma}L(t)^{\eta+\gamma}=c'~.
\end{equation}

Looking for the large time asymptotic regime for which 
$L(T)$ and $A(t)$ (which are assumed to be monotonously increasing) 
become much larger that the constant $c'$, we can solve for $A(t)$ and $L(t)$ as follows.
\begin{itemize}
\item Hence,
\begin{align}
L(t) & = \left[ \frac{1}{2-\alpha-\theta} A(t)^{2-\alpha-\theta} ( \eta+\gamma ) \right]^{1/(\eta+\gamma)} \\
 & = c_2 A(t)^\frac{2-\alpha-\theta}{\eta+\gamma}~.  \label{tttnw}
\end{align}

Plug-in this into equation~(\ref{eq:ODE_system1}) leads to
\begin{equation}
\frac{dA}{dt} = c_2 A(t)^{2-\alpha}~.
\end{equation}

By separating variables and subsequent integration, we get:
\begin{align}
A(t)^{\alpha-2} dA & = c_2  dt~, \\
\frac{1}{\alpha-1} A(t)^{\alpha-1} & =c_2 t + c_2'~, \\
A(t) & = \left[ (1-\alpha) c_2(\frac{-c_2'}{c_2} - t) \right]^{-1/(1-\alpha)} \\
\Leftrightarrow A(t) &  = A_0 (t_c-t)^{-1/\mu}~,  \label{yju6k7k}
\end{align}
with $\mu=1-\alpha$.

\item Similar, we find the solution for $L(t)$:
\begin{align}
A(t) & = \left[ \frac{1}{\eta+\gamma} L(t)^{\eta+\gamma} \right]^{1/(2-\alpha-\theta)} \\
 & = c_3 L(t)^\frac{\eta+\gamma}{2-\alpha-\theta}~.
\end{align}

Plug-in this into equation~(\ref{eq:ODE_system2}) leads to
\begin{align}
\frac{dL}{dt} & = c_3' L(t)^{{\frac{(\eta+\gamma)(1-\alpha)}{2-\alpha-\theta}+1}} \\
& =: c_3' L(t)^{\kappa+1} \quad \text{where}~\kappa  := \frac{(\eta+\gamma)(1-\alpha)}{2-\alpha-\theta}~.
\end{align}
As before, we separate variables and integrate
\begin{align}
L(t)^{-\kappa-1} dL & = c_3' dt~, \\
\frac{1}{-\kappa} L(t)^{-\kappa} & = c_3' t + c_3''~, \\
L(t) & = \left[ \kappa c_3'(\frac{-c_3''}{c_3'}-t) \right]^{-1/\kappa} \\
\Leftrightarrow L(t) & = L_0(t_c-t)^{-1/\kappa}~,
\end{align}
with $\kappa= \frac{\eta+\gamma}{2-\alpha-\theta} (1-\alpha)$.
\end{itemize}
Of course, the solution for $L(t)$ could be directly obtained using (\ref{tttnw})
and (\ref{yju6k7k}), and reciprocally.

For a general mathematical rigorous theory of ordinary differential equations
exhibiting finite-time singular behaviors, see \cite{Goriely2000}.

%%%%%%%%%%%%%%%%%%%%%%%%%%%%%%%%%%%%%%%
\section{Calculation of the exponent $\varphi$}

Let us give some intermediate steps towards the solution of equation~(\ref{eq:solution}).
\begin{align}
  \frac{Y(t)}{A(t)^\xi} &  =^{(\ref{eq:CobbDouglas})}  \frac{K(t)^\alpha(A(t)L(t))^{1-\alpha}}{A(t)^\xi} \\
    & =^{(\ref{eq:AssumptionKremer})}  L(t)A(t)^{1-\alpha-\xi} \\
    & =^{(\ref{eq:solutionTechnology}, \ref{eq:solutionLabour})}  L_0(t_c-t)^{-1/\kappa} \left[ A_0(t_c-t)^{-1/\mu} \right]^{1-\alpha-\xi}\\
    & = L_0 A_0 (t_c-t)^{-1/\kappa - (1-\alpha-\xi)/\mu} \\
    & =^{!}  C_0(t_c-t)^{-1/\varphi}~.
\end{align}

Hence,
\begin{align}
  \varphi & = \frac{1}{1/\kappa-\xi /  \mu+1}~.
%    & =^{(\ref{eq:SolutionDelta}, \ref{eq:SolutionKappa})}  1/(\frac{{\eta+\gamma}{2-\theta-\alpha}(1-\alpha)1/(1-\alpha)-1+\xi/(1-\alpha))
\end{align}

%%%%%%%%%%%%%%%%%%%%%%%%%%%%%%%%%%%%%%%%%%%%%%%%%%%%%%%%%%%%%%%%%%%%%%%%%%%%%%
%\section{Further plots}

%%+++++++++++++++++++++++++++++++++++++++++++++++++++++++++++++++++++++++++++++
% population_percent
%%+++++++++++++++++++++++++++++++++++++++++++++++++++++++++++++++++++++++++++++

%%+++++++++++++++++++++++++++++++++++++++++++++++++++++++++++++++++++++++++++++
% co2_overview
%%+++++++++++++++++++++++++++++++++++++++++++++++++++++++++++++++++++++++++++++

%%+++++++++++++++++++++++++++++++++++++++++++++++++++++++++++++++++++++++++++++
% co2_all
%%+++++++++++++++++++++++++++++++++++++++++++++++++++++++++++++++++++++++++++++

%% population plot (early):

%% population plot (late):

%% co2 plot (early):

%% co2 plot (late):

%% co2_special1

%\renewcommand{\thefigure}{SI \arabic{figure}}
%\setcounter{figure}{0}  % reset counter 

%% linear/exp/power-law comparison

%% population growth in barplot

%% co2 overview:

%% co2 plot (all):

%%+++++++++++++++++++++++++++++++++++++++++++++++++++++++++++++++++++++++++++++
%% sweave clean up:
%%+++++++++++++++++++++++++++++++++++++++++++++++++++++++++++++++++++++++++++++

%%%%%%%%%%%%%%%%%%%%%%%%%%%%%%%%%%%%%%%%%%%%%%%%%%%%%%%%%%%%%%%%%%%%%%%%%%%%%%%
% hack to show figure numbers:
%\label{fig:population_fit_early}
%\label{fig:population_late}
%\label{fig:co2_fit_early}
%\label{fig:co2_fit_late}
%\label{fig:co2_special1}
%\label{fig:comparison}
%\label{fig:population_percent}
%\label{fig:ODE_system}
%\label{fig:co2_overview}
%\label{fig:co2_all}
%%%%%%%%%%%%%%%%%%%%%%%%%%%%%%%%%%%%%%%%%%%%%%%%%%%%%%%%%%%%%%%%%%%%%%%%%%%%%%%
\end{document}